\documentstyle[12pt,epsfig]{article}
\author{Juli\'an Candia$^{a}$ and Ezequiel V. Albano$^{b}$\\{}\\
$^a${\small\it Departamento de F\'{\i}sica, UNLP, 
CC67, 1900 La Plata, Argentina}\\
$^b${\small\it Instituto de Investigaciones Fisicoqu\'{\i}micas
Te\'{o}ricas y Aplicadas}\\{\small\it (INIFTA), UNLP, CONICET, 
Suc.4, CC16, 1900 La Plata, Argentina}}

\title{Monte Carlo simulation of the 
irreversible growth of magnetic thin films}
\begin{document}
\maketitle

\begin{abstract}
\indent The growth of magnetic films
with ferromagnetic interactions between
nearest-neighbor spins is studied in $(d+1)-$dimensional rectangular
geometries for $d = 1,2$. 
Magnetic films are grown irreversibly
by adding spins at the boundaries of the growing interface. 
The orientation of the added spins
depends on both the energetic interaction with already deposited spins and the 
temperature, through a Boltzmann factor. 
At low temperatures thin films, of thickness $L$, are constituted
by a sequence of well ordered magnetic domains. Spins belonging
to each domain, of average length $l_{D} \gg L$, have mostly the same
orientation, but consecutive domains have opposite magnetization.
Such kind of ``spontaneous magnetization reversal'' during the
growth process has a short characteristic length $l_{R}$, such
that $l_{D} \gg l_{R} \sim L$.
At higher temperatures, a transition between ordered and 
disordered states is also observed. The emerging behavior
is compared to that of the equilibrium Ising model. 
\end{abstract}

\section{Introduction}

The preparation and characterization of magnetic nanowires and 
films is of great interest for the development of advanced
microelectronic devices. Therefore, the study of the 
behavior of magnetic materials in confined geometries,
e.g. thin films, has attracted both experimental, see e.g.
 \cite{iron,nw1,hong,shen,tsay,nw2,w110}, and theoretical, 
see e.g. \cite{eva,lan,kar,ou,reis}, attention. 
From the theoretical point of view, most
of the work has been devoted to the study of equilibrium properties of
thin magnetic films \cite{eva,lan,kar,ou,reis}. 
In contrast, the aim of this work is to study
the properties of thin magnetic film growth under
far-from-equilibrium conditions. Within this context, a
useful model for the study of the growth of magnetic
materials is the so-called Magnetic Eden Model (MEM) \cite{mem1},
based on the well known Eden model \cite{ede}. The latter has become an
archetypical growth model due to both its simplicity and
interesting properties. Eden growth starts from a single
particle called the seed. 
One then proceeds to add a new 
particle on a randomly chosen unoccupied site in the
immediate neighborhood (the perimeter) of the seed.
The growth process then continues by
randomly adding new particles to the perimeter of the
previously formed cluster.
Although this simple rule leads to the growth of
compact clusters filling the Euclidean space, 
the self-affinity that characterizes the behavior
of the growing interface is of much interest 
(see e.g. \cite{bar,shl1}). 
The MEM, originally motivated 
by the study of the structural properties of magnetically 
textured materials,
introduces an additional degree of freedom to the Eden 
model, namely the spin of the added particles \cite{mem1}.
More recently, the Eden growth of clusters of charged 
particles has also been studied \cite{qe}. 
 
Considering the MEM with spins having two possible 
orientations (up and down), one can start the growth of the spin cluster
from a single seed having a predetermined orientation, e.g.
a spin up seed, placed at the center of the two-dimensional square lattice,
whose sites are labelled by their rectangular coordinates $(i,j)$.
Then, the MEM's growth process 
consists in adding further spins to the growing cluster
taking into account the corresponding interaction energies. By analogy
to the classical Ising model \cite{ising} 
one takes $J$ as the coupling constant between
nearest-neighbor (NN) spins $S_{ij}$ and the energy $E$ given by

\begin{equation}
E = - \frac{J}{2} \sum_
{ \langle ij,i^{'}j^{'} \rangle} S_{ij}S_{i^{'}j^{'}} ,   
\end{equation}
  
\noindent where  $\langle ij,i^{'}j^{'} \rangle$ means 
that the summation is 
taken over occupied NN sites. As we are concerned with spin
$- \frac{1}{2}$ particles,
the spins can assume two values, namely $S_{ij}= \pm 1$.

It is worth mentioning that, while previous studies of the 
MEM were mainly devoted to determine the
lacunarity exponent and the fractal dimension of the set of parallel
oriented spins \cite{mem1}, 
the aim of the present work is to study the growth of
MEM films using extensive Monte Carlo simulations. 
In order to simulate thin film growth, 
our study is performed in confined (stripped) geometries which resemble
recent experiments where the growth of 
quasi-one-dimensional strips of Fe on a Cu(111) vicinal surface
\cite{iron} and Fe on a W(110) stepped substratum \cite{w110}  
have been performed.  
Also, in a related context, the
study of the growth of metallic multilayers have shown a rich 
variety of new physical phenomena. Particularly, the growth
of magnetic layers of Ni and Co separated by a Cu spacer
layer has recently been studied \cite{cobre}.

Another goal of the present work is to compare the 
results obtained for the MEM with the 
well known behavior of the equilibrium Ising model \cite{ising,Wu}, 
an archetypical model in the study of
phase transitions in equilibrium magnetic systems. 
The Ising Hamiltonian $(\bf{H})$ is given by

\begin{equation}
{\bf H} = - \frac{J}{2} \sum_
{ \langle ij,i^{'}j^{'} \rangle} S_{ij}S_{i^{'}j^{'}} \ \ ,   
\end{equation}

\noindent where $\langle ij,i^{'}j^{'} \rangle$ means 
that the summation runs over all NN sites,
$S_{ij} = \pm 1$ is the state of the spin at the site
of coordinates $(i,j)$ and $J$ is the 
coupling constant ( $J > 0$ for the ferromagnetic case).
It should be pointed out that, in spite of the fact that
Eq.(1) and Eq.(2) are similar, 
the MEM describes the irreversible
growth of a magnetic material while the Ising model is suitable for 
the study of a magnetic system under equilibrium conditions,
and hence these two models 
operate under extremely different conditions.    

This paper is organized as follows: in Section 2
we give details on the simulation method, Section 3 is devoted
to the presentation and discussion of the results, while 
our conclusions are finally stated in Section 4.

\section{The Monte Carlo simulation method}

The MEM in $(1 + 1)-$dimensions is studied in the square lattice using a
rectangular geometry $L \times M$ with $M \gg L$. 
The location of each site on the
lattice is specified through its rectangular coordinates $(i,j)$,
($1 \leq i \leq M$, $1 \leq j \leq L)$.
The starting seed for the growing cluster is a column
of parallel oriented spins placed at $i=1$. As described in the
foregoing section, the MEM growth occurs by selectively gluing
spins at perimeter sites. It should be noticed that previous
studies of the MEM were performed using a single spin seed
placed at the center of the sample  \cite{mem1}. 
In this way, previous simulations that followed this
approach were restricted to rather modest cluster sizes, i.e.
containing up to 8000 spins \cite{mem1}. In contrast, the
rectangular stripped geometry used in this work is suitable for the 
simulation of the growth of magnetic films and it also 
has significant technical advantages. Indeed, 
when the growing film interface is
close to reach the limit of the sample ($i=M$) one simply
computes the relevant properties of the irreversibly frozen
film's bulk (in the region where the growing process has
definitively stopped), and subsequently applies an algorithm such that
the interface is shifted towards the lowest possible $i-$coordinate
(while, at the same time, the useless frozen bulk is erased). By
repeatedly applying this procedure the growth process is not
limited by the $M-$value of the lattice.
In the present
work films having up to $10^{9}$ spins have been typically grown.
The described procedure can straightforwardly be extended 
to higher dimensions. In fact, we have also studied the MEM 
in $(2+1)-$dimensions employing
a $L \times L \times M$ geometry ($M \gg L$).
Each site on the lattice is now identified
through the rectangular coordinates $(i,j,k)$, 
($1 \leq i \leq M$, $1 \leq j,k \leq L$), 
and the starting seed for the growing film is taken to be
a plane of $L \times L$ parallel oriented spins placed at $i=1$.
The cutting-and-shifting algorithm is in this case also suitable
in order to allow the growing film to acquire particles
beyond the $i=M$ limit.

As already mentioned, the growth process of a MEM film 
consists in adding further spins to the growing film
taking into account the corresponding interaction energies
given by equation (1). A spin is added to the film with a probability
proportional to the Boltzmann factor $\exp \left(\frac{-\Delta E}{k_BT}
\right)$, 
where $\Delta E$ is the total energy change involved. 
It should be noted that at each step all sites of the growing
perimeter are considered and the probabilities of adding up
and down spins to them have to be evaluated. 
After proper normalization
of the probabilities the growing site and the orientation of the
spin are determined through a pseudo-random number generator.
Throughout this work we set the Boltzmann constant equal
to unity ($k_{B} \equiv 1$), we
consider $J > 0$
(i.e., the ferromagnetic case) and we take the
absolute temperature $T$ measured in units of $J$.

\section{Results and discussion}

Magnetic Eden films grown on a stripped geometry of
finite linear dimension $L$ at sufficiently low temperatures
show an intriguing behavior that we call 
spontaneous magnetization reversal.
In fact, we have observed that long clusters are constituted
by a sequence of well ordered magnetic domains
of average length $l_{D} \gg L$.
\begin{figure}
\centerline{{\epsfxsize=3.5in \epsfysize=1.6in \epsfbox{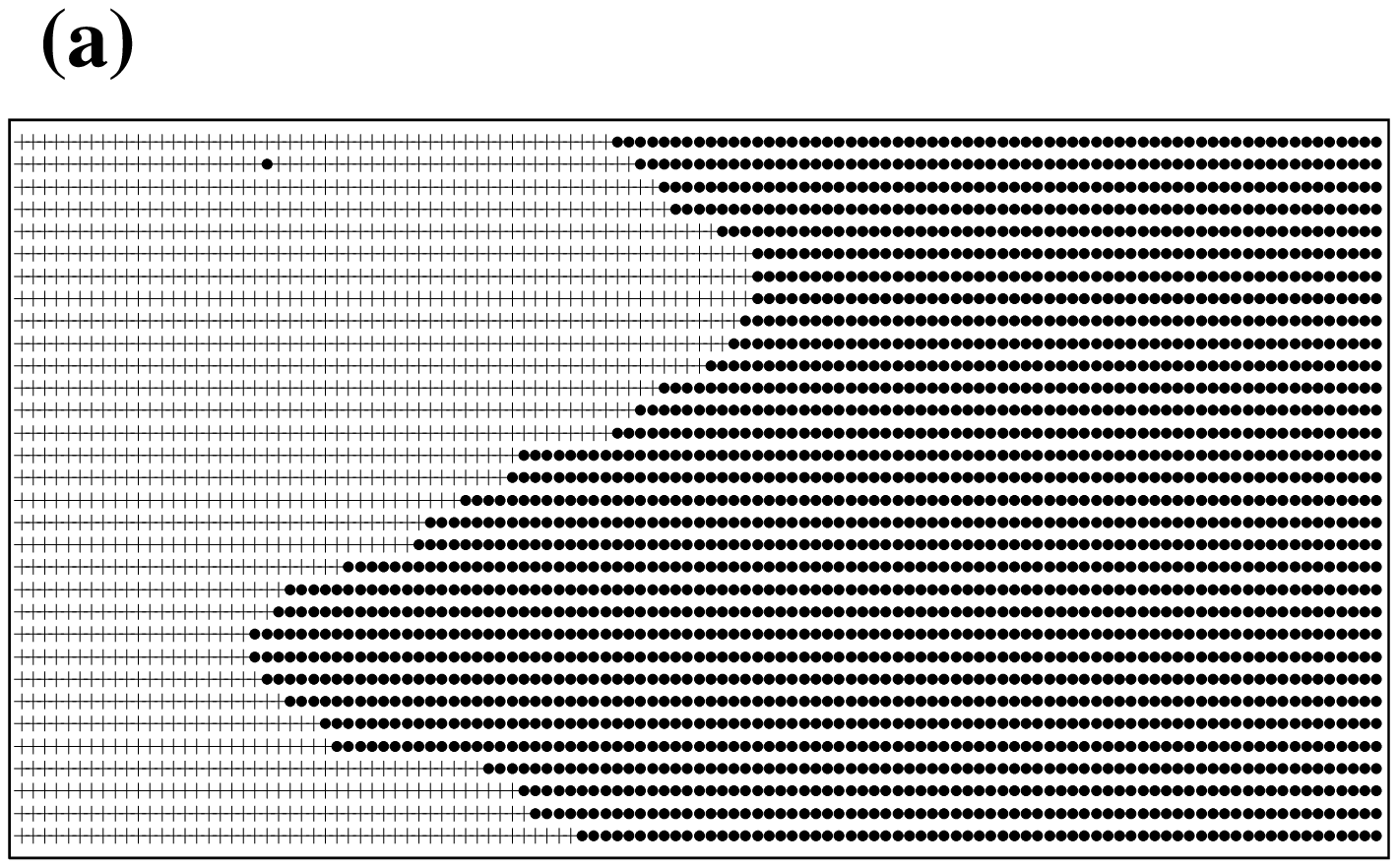}}}
\centerline{{\epsfxsize=3.8in \epsfysize=2.6in \epsfbox{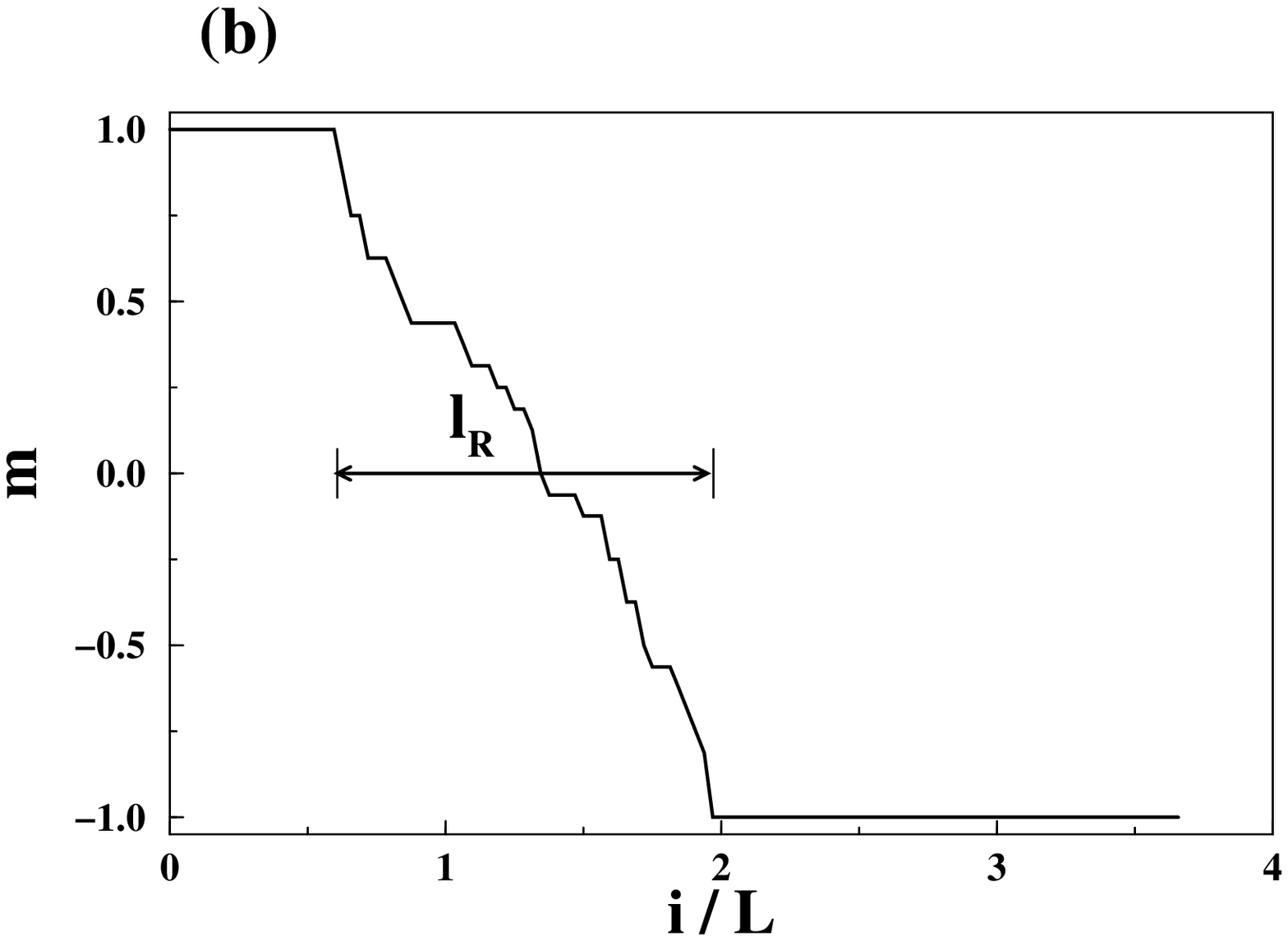}}}
\caption{Spontaneous magnetization reversal observed for $L=32$ and $T=0.26$
in the $(1 + 1)-$dimensional magnetic film.
(a) Snapshot configuration that shows the collective orientation
change: the left (right) domain is constituted by up (down)
oriented spins. The snapshot corresponds to the bulk of the
sample and the growing interface is not shown. 
(b) Magnetization profile associated to the upper
configuration. The characteristic length for the occurrence of the
magnetization reversal, $l_R$, is of the order of the lattice width,
as marked in the figure.}
\label{fig1}
\end{figure}
Figure 1(a) shows a snapshot configuration of the $(1+1)-$dimensional MEM 
where the phenomenon
of spontaneous magnetization reversal can be recognized. Here
the reversal occurring between a domain of spins up (on the
left side) and other one constituted by spins down (on the
right), as well as  the interface between both domains, can be clearly
observed. It should be noted that the well known phenomenon of
field induced magnetization reversal in thin films \cite{a,b}
is quite different from the spontaneous reversal reported here. In the
present study the reversal occurs during the growth process and in the
absence of any applied magnetic field. The reported phenomenon is
essentially due to the small size of the thin film and it 
becomes irrelevant in the thermodynamic limit. Within our 
best knowledge this theoretical prediction has not yet been observed 
experimentally. However, it will certainly be very interesting to design
and carry out suitable experiments in order to observe this phenomenon.

Let $l_{R}$ be the characteristic length for the occurrence
of the spontaneous magnetization reversal. 
Since $l_{R}\sim  L$, we then conclude that the phenomenon
has two characteristic length scales, namely $l_{D}$ and $l_{R}$,
such that $l_{D} \gg l_{R} \sim L$. Figure 1(b) shows the magnetization
profile that corresponds to the spontaneous magnetization reversal shown in
figure 1(a), where

\begin{equation}
m(i,L,T) = \frac{1}{L} \sum_{j = 1}^{L} S_{ij}                
\end{equation}

\noindent is the mean column magnetization at the distance
$i-1$ from the seed, for a system of linear dimension $L$ at
temperature $T$. In figure 1(b) one can clearly observe how abruptly
the magnetization drops from $m=+1$ to $m=-1$ within
a characteristic length $l_{R}$ of the order of $L$.

\begin{figure}
\centerline{{\epsfxsize=3.9in \epsfysize=2.7in \epsfbox{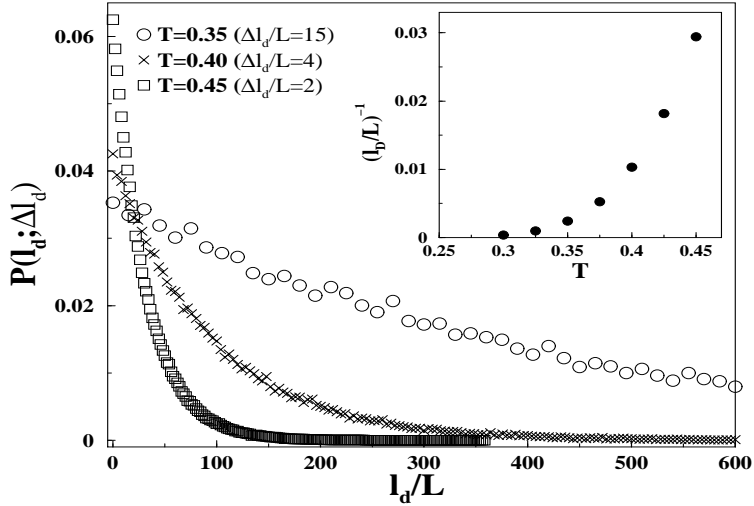}}}
\caption{Plots of $P(l_d;\Delta l_d)$ versus $l_d/L$ for $L=16$ and
different values of temperature and interval width, as indicated. 
The inset shows a plot of $(l_D/L)^{-1}$ versus $T$, also for a 
lattice of side $L=16$. As expected, increasingly long domains tend
to show up at lower temperatures.}
\label{fig2}
\end{figure}

\begin{figure}
\centerline{{\epsfxsize=3.8in \epsfysize=2.5in \epsfbox{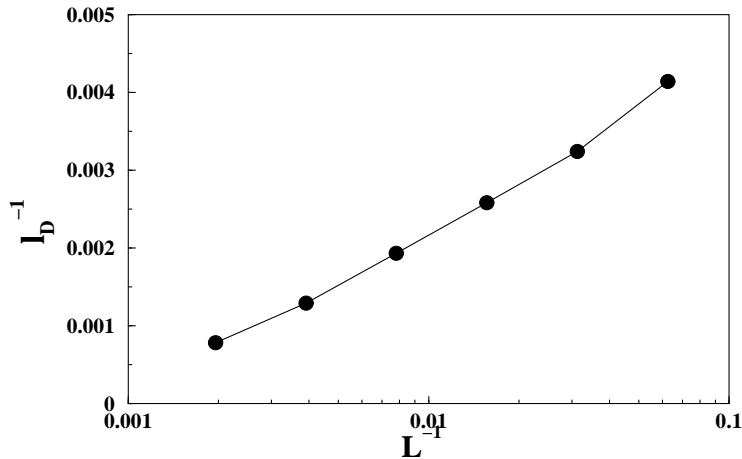}}}
\caption{Log-linear plot of $l_D^{-1}$ versus $L^{-1}$ for the temperature  
$T=0.5$ and lattice sizes in the range $16 \leq L \leq 512$. The line
is a guide to the eye, showing that the characteristic domain 
length diverges in the thermodynamic
limit ($L \rightarrow \infty$).}
\label{fig3}
\end{figure}

In order to investigate the dependence of the characteristic 
domain length $l_D$ on $L$ and $T$, let us define 
$P(l_d;\Delta l_d)$ as the probability for the 
formation of a domain of length between $l_d$ and $l_d+\Delta l_d$. 
Clearly, the average domain length $l_D$ mentioned above is 
the average value of $l_d$ taken over a sufficiently long 
magnetic film (i.e. $l_D \equiv \langle l_d \rangle$
for $l_F \gg l_D$, where $l_F$ is the film's total length). Figure 2 shows
plots of $P(l_d;\Delta l_d)$ versus $l_d/L$ for a fixed lattice size and
different values of temperature. As expected, increasingly long domains tend
to show up at lower temperatures. This behavior can be also observed in the 
plot of $(l_D/L)^{-1}$ versus $T$ shown in the inset of figure 2. 
Figure 3 shows
a log-linear plot of $l_D^{-1}$ versus $L^{-1}$ for a fixed temperature  
and lattice sizes in the range $16 \leq L \leq 512$. From this figure it 
turns clearly out that the average domain length diverges as we increase the
lattice size towards the thermodynamic limit. Therefore, as early
anticipate the phenomenon of magnetization reversal is a finite
size effect releveant for the growth of magnetig films in confined
geometries. 

At this point it is useful to perform a comparison between the results obtained
with the MEM and the well known behavior of the Ising model 
analyzing the interplay between broken symmetry and finite-size 
effects at thermal phase transitions. 
In ordinary thermally driven phase transitions, the system
changes from a disordered state at high temperatures to a
spontaneously ordered state at temperatures below some critical
value $T_{c}$ where a second-order phase transition takes place.
Regarding the equilibrium Ising model as the archetypical example,
one has that, in the absence of an externally applied magnetic
field ($H=0$), the low temperature ordered phase is a state with
non-vanishing spontaneous magnetization  ($ \pm M_{sp}$, the sign 
depending on the initial state).
This spontaneous symmetry breaking is possible
in the thermodynamic limit only. In fact,
it is found that the magnetization $M$ of a finite sample formed
by $N$ particles, defined by

\begin{equation}
M(T,H=0) = \frac{1}{N} \sum_{i = 1}^{N} S_{i}(T,H=0)  \ \ ,
\end{equation}

\noindent can pass with a finite probability from a value near
$+M_{sp}$ to another
near $-M_{sp}$, as well as in the opposite direction. Consequently,
the magnetization of a finite system, averaged over a
sufficiently large observation time, vanishes at every positive temperature,
irrespective of the (finite) size of the sample.
The equation $M(T,H=0) \approx 0$ holds if the observation time ($t_{obs}$)
becomes larger than the ergodic time ($t_{erg}$), which is defined as
the time needed to observe the system passing from $\pm M_{sp}$ 
to $\mp M_{sp}$. 
Increasing the size of the sample the ergodic time increases too,
such that in the thermodynamic limit ergodicity is
broken due to the divergence of the ergodic time, yielding
broken symmetry.
Since Monte Carlo simulations are restricted
to finite samples, the standard procedure to avoid the problems
treated in the foregoing discussion is to consider the root mean
square (or the absolute) magnetization as an appropriate order
parameter \cite{kuku}.
Turning back to the MEM, we find  that the phenomenon of magnetization
reversal (as shown in figure 1(a)) causes the
magnetization of the whole film to vanish at every non-zero
temperature, provided that the film's length $l_{F}$ (which plays the 
role of  $t_{obs}$) is much larger than $l_{D}$ (which plays the role
of $t_{erg}$ ).
Therefore, as in the case of the Ising model \cite{kuku}, 
in order to overcome shortcomings derived from the 
finite-size nature of Monte Carlo
simulations we have measured the 
mean absolute column magnetization, given by

\begin{equation}
|m(i,L,T)| = \frac{1}{L} |\sum_{j = 1}^{L} S_{ij}| \ \ .                             
\end{equation}

In the stripped geometry used in this work the bias introduced
by the linear seed (a starting column made up entirely of up spins)
can be studied in plots of $|m(i,L,T)|$ versus $i$.
It is found that $|m(i;L,T)|$ exhibits
a transient growing period with a characteristic length 
of order $L$, followed by the attainment of a stationary regime. 
In addition, using several randomly generated seeds
we could also establish that the system evolves into
a given stationary state independently of the seed employed.
Thus, the spin-up linear seed can be used throughout 
without loss of generality.
This behavior has been
observed throughout the range of interest 
studied in the present work, i.e. $16 \leq L \leq 1024$ and
$0.2 \leq T \leq \infty$.
So, the influence exerted on the spin system by the seed can
be easily recognized and eliminated from our results just by
disregarding the first $l_{Tr} = N.L$ columns, 
with $N$ ranging between 10 and 50. 
The given procedure of column averaging out from the
transient region represents a significant advantage
of the stripped geometry used for the
simulation of the MEM, in addition to that
already mentioned (see section II). In fact, when a single seed 
at the center of the sample is used, the definition
of the average magnetization of the whole cluster 
is strongly biased by the cluster's
kernel orientation at the early stages of the growing process.
Hence, it turns very difficult in this case to disentangle the
stationary regime from the transient region.
Moreover, film growth on planar substrata has the 
advantage that it can be implemented experimentally,
e.g. via vapor deposition in vacuum, chemical deposition, 
electrodeposition, etc.   

\begin{figure}
\centerline{{\epsfxsize=3.9in \epsfysize=2.7in \epsfbox{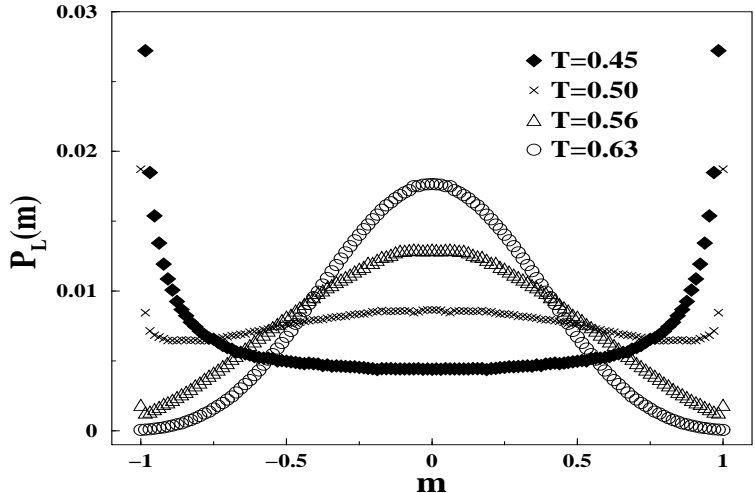}}}
\caption{Data corresponding to the $(1 + 1)-$dimensional MEM: 
plots of the probability distribution of the mean
column magnetization $P_{L}(m)$ versus $m$ for the fixed lattice
width $L=128$ and different temperatures, as indicated in
the figure.
The sharp peaks at $m = \pm 1$ for $T=0.45$ have been
truncated, in order to allow a detailed observation of the plots
corresponding to higher temperatures.
This behavior resembles that of the one-dimensional
Ising model.}
\label{fig4}
\end{figure}

The mean column magnetization given by equation (3)
is a fluctuating quantity that can assume $L+1$ values. 
Then, for given values of both $L$ and $T$,
the probability distribution of the mean column
magnetization $(P_{L}(m))$ 
can straightforwardly be evaluated, since it represents the
normalized histogram of $m$ taken over a sufficiently large number
of columns in the stationary region \cite{alfa,beta,gamma}.
In the thermodynamic limit (lattice size going to infinite)
the probability distribution $(P_{\infty}(m))$
of the order parameter of an equilibrium system
at criticality is universal (up to rescaling of the 
order parameter) and thus it contains
very useful and interesting information
on the universality class of the system \cite{pd1,pd2,pd3}. 
For example, $P_{L}(m)$ contains information about all
momenta of the order parameter $m$, including 
universal ratios such as the Binder cumulant \cite{pd1}. 
Figure 4 shows the thermal dependence of $P_{L}(m)$
for a fixed lattice size ($L=128$ in the present example)
as obtained for the $(1 + 1)-$dimensional MEM.
We can observe that at high temperatures $P_{L}(m)$ is a Gaussian
centered at $m=0$. As the temperature gets lowered, the distribution
broadens and develops two peaks at $m=1$ and $m=-1$.
Further decreasing the temperature
causes these peaks to become dominant
while the distribution turns distinctly non-Gaussian, exhibiting a
minimum just at $m=0$.
It should be pointed out that the emergence of the maxima
at $m= \pm 1$ is quite abrupt.
This behavior reminds us the order parameter probability
distribution characteristic of the one dimensional Ising model.
In fact, for the well studied $d-$dimensional Ising model \cite{gamma,lali}, 
we know that
for $T > T_{c}$, $P_{L}(M)$\cite{note} is a 
Gaussian centered at $M=0$, given by

\begin{equation}
P_{L}(M) \propto \exp \left({-M^{2}L^{d}} \over {2T \chi} \right) \ \    ,    
\end{equation}

\noindent where the susceptibility $\chi$ is related to
order parameter fluctuations by

\begin{equation}
\chi = \frac {L^{d}}{T} \left(\langle M^{2} \rangle - \langle M \rangle^{2} \right)  .                 
\end{equation} 

Decreasing temperature the order parameter probability 
distribution broadens, it becomes non-Gaussian, and near $T_{c}$ it
splits into two peaks that get the more separated the lower the
temperature. For $T < T_{c}$ and linear dimensions $L$ much larger
than the correlation length $\xi $ of order parameter fluctuations,
one may approximate $P_{L}(M)$ near the peaks by a double-Gaussian
distribution, i.e.

\begin{equation}
P_{L}(M) \propto \exp \left( {-(M - M_{sp})^{2}L^{d}} \over {2T \chi} \right) +
\exp \left( {-(M + M_{sp})^{2}L^{d}} \over {2T \chi} \right) \ \ ,   
\end{equation}                          

\noindent where $M_{sp}$ is the spontaneous magnetization, while
the susceptibility $\chi$ is now given by 

\begin{equation}         
\chi = \frac{L^{d}}{T} \left(\langle M^{2}\rangle - \langle |M| \rangle ^{2} \right) \ \ .
\end{equation}

\noindent From equation (6) it turns out that 
the Gaussian squared width $\sigma^{2}$
associated with high temperature distributions is very close to
the 2nd moment of the order parameter, i.e.

\begin{equation}  
\sigma^{2} \approx \langle M^{2} \rangle \ \ .
\end{equation}  

\noindent It should be noticed that this equation is a straightforward
consequence of the Gaussian shape of the order parameter probability
distribution and, thus, it holds for the MEM as well. 
>From the well known one-dimensional exact solution for a
chain of $L$ spins \cite{delta} one can establish the relationship

\begin{equation}   
\chi = \frac {1}{T} \exp (2/T) \ \ ;
\end{equation}   

\noindent then, equations (7) and (11) lead us to 

\begin{equation}  
\langle M^{2} \rangle = \frac {1}{L} \exp (2/T) \ \          
\end{equation}  

\noindent (where it has been taken into account that
$\langle M \rangle = 0$ due to finite-size effects, irrespective
of temperature).
From equations (10) and (12) we can see that the high
temperature Gaussian probability distribution broadens exponentially
as $T$ gets lowered, until it develops delta-like peaks at $M= \pm 1$ 
as a consequence
of a boundary effect on the widely extended distribution.
It should be noted that for $d \geq 2$ this phenomenon 
is prevented by the finite
critical temperature which splits the Gaussian, as implied by equation (8).

\begin{figure}
\centerline{{\epsfxsize=3.9in \epsfysize=2.7in \epsfbox{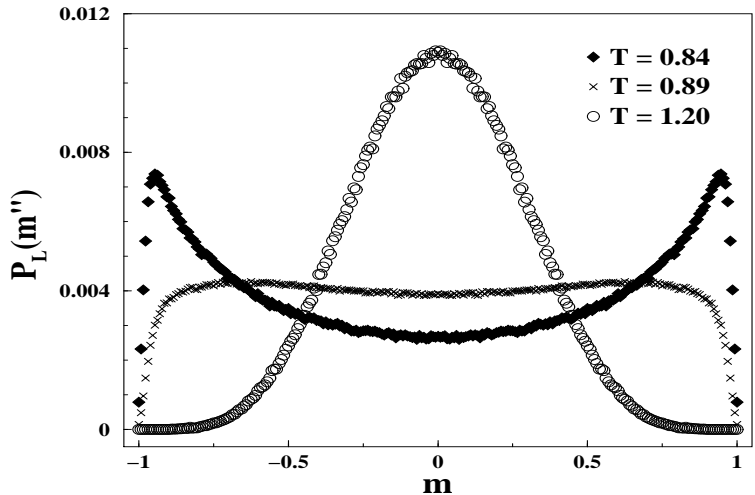}}}
\caption{Data corresponding to the $(2 + 1)-$dimensional MEM:
plots of the probability distribution 
$P_{L}(m^{''})$ versus $m^{''}$ for the fixed lattice
size $L=16$ and different temperatures, as indicated in
the figure. 
The occurrence of two maxima located at 
$m^{''} = \pm M_{sp}$ (for a given value of $M_{sp}$
such that $0 < M_{sp} < 1$)
is the hallmark of a thermal continuous phase transition
that takes place at a finite critical temperature.}
\label{fig5}
\end{figure}

Figure 5 shows the thermal evolution of the probability
distribution as obtained for the $(2 + 1)-$dimensional MEM, 
using a lattice of side $L=16$.
For high temperatures, the probability distribution
corresponds to a Gaussian centered at $m^{''}=0$. At lower
temperatures we observe the onset of two
maxima located at $m^{''} = \pm M_{sp}$ $(0 < M_{sp} < 1)$,
which become sharper and approach 
$m^{''} = \pm 1$ as $T$ is gradually decreased.
These low-temperature probability distributions 
clearly reflect the occurrence of the
magnetization reversal effect already discussed for
the case of $(1 + 1)-$dimensional magnetic films.

\begin{figure}
\centerline{{\epsfxsize=3.8in \epsfysize=2.5in \epsfbox{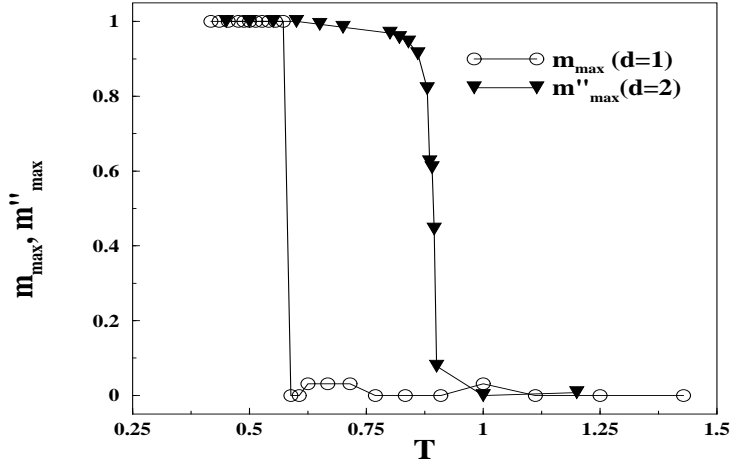}}}
\caption{Plots showing
the location of the maximum of the probability distribution
as a function of temperature for both 
$(d+1)-$dimensional MEM models ($d=1,2$). The lines are guides to the eye.
The smooth transition
for d=2 constitutes another evidence of the
non-zero critical temperature associated with the $(2+1)-$MEM.}
\label{fig6}
\end{figure}

Figure 6 shows
the location of the maximum of the probability distribution
as a function of temperature for both 
$(d+1)$-dimensional MEM models (with  $d=1,2$)
where we consider only maxima located at $m,m^{''} \geq 0$,
since the distributions are symmetric around $m=m^{''}=0$.
After inspection of figure 6, 
it becomes apparent the different qualitative behaviors of both systems. 
In fact, while for the $d=2$ case we observe a smooth
transition from the $m^{''}_{max}=0$ value characteristic of high temperatures
to nonzero $m^{''}_{max}$ values that correspond to lower temperatures,
the curve obtained for $d=1$ shows, in contrast, a Heaviside-like jump.
The latter case reflects a behavior which is similar to that observed 
simulating the 
equilibrium Ising model in 1d. 

In order to carry out a quantitative comparison between both models, 
we have also evaluated 
the average absolute magnetization ($\langle |M_{Ising}|  \rangle$) for chains
of the same length $L$ that the columns of the $(1 + 1)-$dimensional
MEM. Figure 7 shows log-linear plots of
$ \langle |M|_{Ising}  \rangle (L,T) -  \langle |m|_{MEM}  \rangle (L,T)$  
versus $L^{-1}$
obtained for two different values of $T$. It becomes evident
that the different $L-$dependent values of the magnetization
are finite size effects observed as a consequence of
the strips used. Such effects vanish in the thermodynamic limit.

\begin{figure}
\centerline{{\epsfxsize=3.8in \epsfysize=2.5in \epsfbox{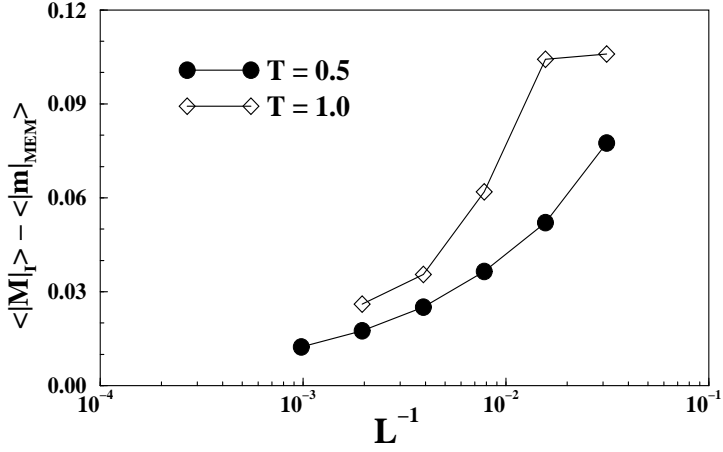}}}
\caption{Comparison of results corresponding to the 
$(1+1)-$MEM and the $d=1$ Ising model:  
log-linear plots of $\langle |M|_{I} \rangle (L,T) - \langle |m|_{MEM}
\rangle (L,T)$  versus $L^{-1}$
for $T=0.5$ and $T=1.0$. 
The lines are guides to the eye.
Hence, differences in the magnetization 
due to  finite-size effects appear to
vanish in the thermodynamic limit.}
\label{fig7}
\end{figure}

In contrast to the $(1 + 1)-$dimensional case, the behavior exhibited by the
$(2+1)$-dimensional MEM 
(e.g., as displayed by figures 5 and 6) is the signature
of a thermal continuous phase transition that takes place at
a finite critical temperature. It should be noticed that this 
transition involves the entire system, that may be either
in the disordered phase or in the ordered one, depending on 
temperature. 
The broken symmetry at a finite critical temperature $T_{c}$ implied by
the thermal continuous phase transition can be explained in terms of the
broken ergodicity that occurs in the system when we tend to
the thermodynamic limit ($L \rightarrow \infty$)
making use of the temperature dependence exhibited by the
order parameter distribution function.
In fact, if we introduce an ergodic length $l_{erg}$
defined by $l_{erg} \equiv l_D$, where $l_D$ is
the characteristic length of MEM's domains,
we can carry out a complete analogy with the Ising model
by associating $l_{erg}$ to $t_{erg}$ (the Ising model ergodic time)
and the above mentioned film's total length $l_{F}$
to the Ising model observation time $t_{obs}$.
In this way, we encounter that
excursions of $m^{''}$ from $m^{''} = +M_{sp}$ to $m^{''} = -M_{sp}$ and
{\it vice versa} occur at
length scales of the order of  $l_{erg}$.
When the film's
total length becomes larger and larger ($l_{F} \gg l_{erg}$) the whole
film's magnetization is averaged to zero.
Furthermore, $l_{erg}$ diverges
as the strip's width becomes larger and larger, 
as shown in figure 3, and again broken
symmetry arises as the consequence of broken ergodicity.

\begin{figure}
\centerline{{\epsfxsize=3.8in \epsfysize=2.5in \epsfbox{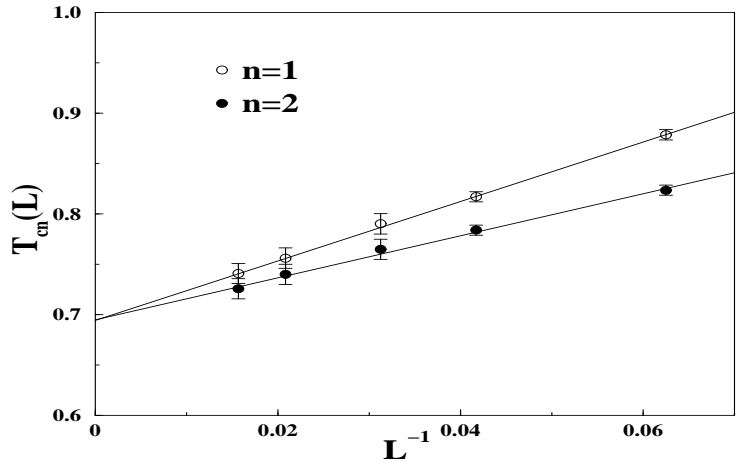}}}
\caption{Plots of the effective finite-size critical temperatures
$T_{cn}(L)$ versus $L^{-1}$ (for $n=1,2$)
corresponding to the $(2+1)-$dimensional magnetic film. 
$T_{c1}(L)$ is defined
as the value that corresponds to 
$\langle |m^{''}| \rangle = 0.5$,
while $T_{c2}(L)$ is the temperature that corresponds to 
the maximum of the susceptibility.
The solid lines show the linear extrapolations that meet at the
critical point given by $T_{c}=0.69 \pm 0.01$.}
\label{fig8}
\end{figure}

It should be noted that as in the case of equilibrium systems, 
in the present  case various ``effective" L-dependent critical 
temperatures can also be defined. In particular, we will define $T_{c1}(L)$ 
as the value that corresponds to 
$\langle |m^{''}| \rangle = 0.5$ for fixed $L$,
and $T_{c2}(L)$ as the temperature that corresponds to 
the maximum of the
susceptibility for a given $L$, assuming that the susceptibility
is related to order parameter fluctuations in the same manner
as for equilibrium systems (as given by equations (7) and (9)).
Then, we should be able to obtain $T_{c}$ from
plots of $T_{cn}(L)$ versus $L^{-1}$  (for $n=1,2$),
as it is shown in figure 8.
Indeed, following this procedure we find that, for $L \rightarrow \infty$, 
both $T_{c1}(L)$ and $T_{c2}(L)$
extrapolate (approximately) to the same value, allowing us 
to evaluate the critical temperature
$T_{c} = 0.69 \pm 0.01 $ in the thermodynamic limit. 
Notice that $T_{c}$ depends on 
both the coordination number
and the topological structure of the lattice.
Furthermore, these properties are not uniquely determined by the
dimensionality. So, the small
value of $T_{c}$ obtained for the MEM, as compared to the Ising model
on the square lattice with four nearest neighbors (NN)
given by $T_{c}^{Ising} = 2.27$ \cite{delta}, reflects the fact that
the effective number of occupied NN sites 
upon deposition of spins in the MEM ($\langle NN_{MEM} \rangle$) should be
$\langle NN_{MEM} \rangle  < 4$. Furthermore, the effective topological 
structure of the MEM compatible with the measured value
of $T_{c}$ remains as an open question. 

\section{Conclusions}

In the present work we have studied the growth of magnetic
films with ferromagnetic interactions between
nearest neighbor spins in a $(d+1)-$dimensional rectangular
geometry (for $d=1,2$), using Monte Carlo simulations.
For both dimensions the phenomenon of 
spontaneous magnetization reversal
is observed at low temperatures. Indeed, MEM films grown
at low temperatures   
are constituted by a sequence of magnetic domains, each of them 
with a well defined magnetization, such that the magnetization of
adjacent domains is antiparallel. 
Further increasing the temperature causes the onset
of disorder in the bulk of the domains. Subsequently,
a rounded effective transition to a fully disordered 
state takes place. These pseudo ``phase transitions'' occur
at film width ($L$) dependent effective critical temperatures.
However, in the thermodynamic limit ($L \rightarrow \infty$),
the $(1 + 1)-$dimensional MEM is not critical (the transition
takes place at $T = 0$), while a true second-order 
phase transition is expected to occur at a finite
temperature ($T_{c} = 0.69  \pm 0.01$) in the $(2 + 1)-$MEM.
The observed behavior is reminiscent to that of the equilibrium Ising 
model, although it should be stressed that the MEM
is a far-from-equilibrium growing system.

The finite size of the films causing magnetization reversal and
the occurrence of effective order-disorder transitions 
may be undesired effects that shall be avoided in the preparation of
high quality magnetic films. However, these shortcomings may
disappear if the film strongly interacts with the substrate
where the actual growing process takes place.
Further studies on the growth of magnetic films in the 
presence of surface magnetic films, that account for 
the interaction with the substrate, are under progress \cite{paper1,paper4}.  

\section*{Acknowledgments} This work was supported  financially  by
CONICET, UNLP, CIC (Bs. As.), ANPCyT and Fundaci\'on Antorchas 
(Argentina) and the Volkswagen Foundation (Germany).


\begin{thebibliography}{99}
\bibitem{note} Notice that according to the definitions given by 
equations (3) and (4) we have used $P_{L}(m)$ and $P_{L}(M)$
for the MEM and the Ising model, respectively.
\end{thebibliography}
\end{document}